\documentclass[a4paper, 10pt, conference]{ieeeconf}      

\title{\LARGE \bf
Survey Paper on Rising Threats of Subverting Privacy Infrastructure
}

\author{Nethra Balasubramanian
}

\begin{document}

\maketitle
\thispagestyle{empty}
\pagestyle{empty}

\begin{abstract}

One of the main challenges faced by a user today is protecting their privacy, especially during widespread surveillance. This led to the development of privacy infrastructures whose main purpose is to guarantee user’s privacy. However, they are being misused by attackers who consider them to be an exploitable resource to perform illegitimate activities. This paper proposes to analyze and assess the threats that occur due to subverting privacy infrastructures. It begins with an outline of critical privacy infrastructures that were developed. An overview of rising threats of subverting the most prominent privacy infrastructure, Tor is presented. Additionally, a brief assessment of the severity of these threats is discussed and the paper concludes by recommending the scope for further research to mitigate the risks due to such threats.

\end{abstract}

\section{INTRODUCTION}

Due to the rapid usage and development of Internet, privacy has become a major concern for individuals, particularly in case of widespread surveillance. With the flexibility and advancement of communication technologies, the open nature of Internet also encouraged illegitimate activities such as prying on user’s personal information, selling information to advertising companies, exploiting  information by government agencies thereby compromising the privacy of user information. Hence, privacy infrastructures emerged as a solution. (ALSABAH and GOLDBERG 2016) Tor, acronym for the Onion Router is an example of such a system. VPN, Proxy servers, Remailers, JAP, I2P are few other examples. These systems are responsible for hiding the identities of communicating parties from their peers, as well as from the adversaries who intend to eavesdrop the traffic flowing through the network. Among them, VPN, proxy servers, TOR and I2P are actively used. (Chakravarty 2014)(Bingdong Li 2013). The success of privacy infrastructures further encouraged the attackers to perform illegal activities by subverting these infrastructures through sophisticated botnets and ransomware. 

Botnets are generally used by attackers to perform malicious activities such as DDoS, personal data theft, spam, bitcoin mining and cyber-espionage. Botnets are centralized overlay networks in which Command-and-Control (C\&C) servers are a single entity responsible for control. Bots connect to these servers to be reachable and a botmaster exists that manages the bots and is aware of the overlay network structure. Major disadvantage in this architecture is that the C\&C servers act as a single point of failure. Hence, if the C\&C servers are compromised, the complete botnet is defeated. Attackers have resorted to this problem by adopting a more resilient unstructured P2P network that has a distributed architecture, or using improved techniques to reduce the detection capability of the C\&C servers. The latter was achieved by taking advantage of the anonymity features of Tor network. Through Tor, a botmaster can access the C\&C servers anonymously and an encrypted routing system is created to avoid detection through traffic analysis. Moreover, Tor provides hidden services in which the client does not need to know the actual address or location of the service, and botmasters can configure the C\&C servers as hidden services. (Casenove and Miraglia 2014).In this paper, we will briefly describe an overview of privacy infrastructures, rising threats of subverting these infrastructures and assess the criticality of such threats based on its capabilities and previous attacks.

\section{Overview of Privacy Infrastructures}

\subsection{Proxy Servers}

A proxy server is used for surfing the web, wherein it acts as an intermediary between the client and server, and client requests are sent with the proxy server’s identity rather than that of the real user thereby hiding the user’s identity. Due to this, the destination server does not log the real IP address and other device information of user. It is not very prominent for anonymous communication because it can be easily attacked or traced. Before the traffic is sent to the proxy, it has to travel through the ISP’s physical network, making it susceptible to man-in-the-middle attacks. An attacker can monitor the traffic between user and proxy server and obtain sensitive information. (Towards an Autonomous System Monitor for Mitigating Correlation Attacks in the Tor Network 2016) (Erdin 2012) (Bingdong Li 2013)

\subsection{Remailers}

Remailers are servers which forwards to destination the messages it receives with embedded instructions, without revealing the sender’s information. In particular, remailers manipulate the addresses in the e-mail headers of the transmitting node with fake addresses. Hence anonymity in e-mails are achieved. Remailers are of three types, Type I (Cypherpunk) that forwards messages to several servers before sending it to the destination and provides multiple layer encryption through PGP public keys, Type II (Mixmaster) which includes padding, message pools and is more resistant to traffic analysis, Type III (MixMinion) which is most secure due to free routing algorithm. (Towards an Autonomous System Monitor for Mitigating Correlation Attacks in the Tor Network 2016) (Erdin 2012) (Bingdong Li 2013)

\subsection{Virtual Private Network (VPN)}

Unlike proxy server, VPN server provides secured communication to the destination by encrypting the traffic between the user and server. The traffic is encrypted irrespective of the type of application being used, thus mitigating the risks of attacks such as man-in-the-middle. Even though it offers secured communication when compared to proxy server, user’s privacy is not guaranteed. There are many cases where VPN providers share user’s information for business profits. (Erdin 2012)

\subsection{Onion Routing}

It is the most prevalent design for low latency communications. The protocol consists of a structure similar to that of an onion, where message is encapsulated with layer by layer encryption. In this mechanism, there is a set of servers called onion routers that is responsible to relay traffic to the clients. Every node has a public and private key. The public key is known to the client in order to set the path of communication. Initially, the client constructs an encrypted tunnel called circuit using public-key cryptography. Once it is set, symmetric key cryptography is used to transfer the data. This protocol has different variations such as Onion Router(TOR) and Invisible Internet Project(I2P) (Erdin 2012)

\subsubsection{Invisible Internet Project (I2P)}

It is an alternative to TOR that supports all regular internet activities such as e-mail, web browsing. Unlike TOR, I2P is more focused on accessing closed darknet rather than the regular Internet websites. I2P offers anonymity services to identity-sensitive applications by building an overlay network of volunteer systems. It is strictly based on UDP, but security can be achieved by including the libraries that allow reliable stream communication on top of the I2P network. It is a closed system in which traffic is routed through other peers and by announcing its peers, it enables new users to join the network. Many applications such as email, peer-to-peer, IRC interact with I2P but it is usually not preferred for low latency applications due to the lack of focus in end-to-end delay. (Bingdong Li 2013) (Tails 2016)

\subsubsection{Onion Router (TOR) }

Tor is a low latency anonymity overlay network that is known to be the most robust privacy tool. It is helpful in preventing user discovery to any entities that are monitoring the network. When packets are transmitted between the user and a destination host, a random path with three nodes are used so that no single node is aware of the complete transmission process thereby providing anonymity. Furthermore, TOR connections are encrypted using TLS protocol. However, connection between exit node and destination is not encrypted and hence exit nodes can observe the content of messages. (Bingdong Li 2013)

\subsubsection*{Functioning of TOR}

The directory authorities in TOR are centralized, trusted servers that track the complete TOR network. Nodes or routers are voluntary computers that are distributed and categorized based on their respective functionality and positions. Before transmitting the data packets to the destination node, it is encrypted several times using public key cryptography. At every relay, one encryption layer is decrypted, which reveals the IP address of the next relay that the packet needs to be forwarded to. This is done until the packet reaches the destination and the reverse process is followed for messages sent to the client. The client’s privacy is maintained,since without the ability to analyze traffic, none of the relays can detect the corresponding message's sender and destination. Through the hidden services feature, services are accessible to anyone with a Tor client without revealing any knowledge about the IP address or location of the server.For additional security, the Tor client does not select same router or relays in the same /16 subnet to be in the same circuit. Tor’s telescopic approach to circuit establishment has two major benefits, one of which is that perfect forward secrecy is achieved due to discarding the session keys when circuit is closed. The other benefit is that, routers need not store the hashes of previously processed onions to prevent replay attacks since replaying one side of a Diffie- Hellman handshake results in a different key which is not of any use to the attacker. (Bingdong Li 2013)

\subsection{Java Anonymous Proxy (JAP)}

JAP is a low latency mix cascades that uses the server provided by volunteers to access the Internet. Several mixes are used to encrypt the packet and maintain the rate of traffic constant in order to avoid rate-based traffic analysis. The program can display all the active mixes from which user can choose the JAP cascades. (Erdin 2012) JAP software is available from many years and is second in popularity after Tor. Commercially, it is called JonDonym.

\section{Rising threats of subverting privacy infrastructure}

Among the privacy infrastructures mentioned, Tor provides highest anonymity. The stealthiness and untraceability features of Tor motivated the attackers to take advantage of this capability and develop Tor-based botnets. By placing the botnet infrastructure in Tor, the Tor hidden services provides anonymous C\&C servers which is difficult to detect or destroy. Furthermore, attacking the server with DDoS seems unfeasible as it would result in the attack of complete Tor network.

\subsection{Tor-based Botnets [Past Work]} 

The idea of hiding botnets in Tor was discussed in 2010, in particular at the DefCon18 (D.Brown 2010) where a C\&C server anonymity implementation using Tor was shown by DannisBrown. Later in 2012, Guarnieri in (Guarnieri 2012) detected and analyzed the first Tor-based botnet. The botnet was a modified version of Zeus consisting of DDoS, bitcoin mining and credential theft capabilities. The malware comprised of Zeus bot, Tor client, GMinerbitcoin mining tool, and few libraries for GPU hash cracking. The bots ran inside hidden services and all C\&C communication was within the Tor network. The botmaster tried to reduce the traceability by avoiding the use of exit nodes, and used IRC protocol to communicate and issue commands to the bots. The botmaster also made the bots act as relays thereby exploiting and enhancing the Tor network simultaneously. In 2013, due to a post on the Tor mailing list, Tor's network usage and the number of users accessing grew rapidly. Researchers couldn’t explain the reason at first but on analysis came to conclusion that it was due to a large botnet that suddenly switched to Tor. The botnet used centralized structure with HTTP protocol and a preconfigured earlier version of Tor to connect to the network. The significant increase in the amount of Tor communication that was being established through relays resulted in the reduction of Tor system’s responsiveness. (Casenove and Miraglia 2014)

Similarly,in the last few years, various types of botnets were found in (Constantin 2012) (Dunn 2013) (Gottesman 2014)that made use of Tor infrastructure. It provided a hideout for malware by deploying the command and control server as a hidden service with specific onion address that the other bots are configured with. Such botnets and referencing the ones mentioned earlier as well, result in significant degradation in the performance of Tor. One other example is the spike caused by Mevade/Sefnit botnet in 2013. The spike was close to 600\% increase in the number of clients, causing a network overload through C\&C descriptors and creating circuit requests. (ALSABAH and GOLDBERG 2016)

\subsection{OnionBots [Future Threat]}

According to (Sanatinia and Noubir 2015), onion bots are believed to be the next generation of resilient and stealthy botnets. It uses privacy infrastructures such as Tor to stay undetected and decouple itself from the infected hosts. Since onion bots are different from the peer-to-peer botnets, the existing solutions that are used for the peer-to-peer bot are not applicable to onion bots. The design is also resilient to the current mitigation and analysis techniques such as botnet mapping, hijacking and assessing the botnet size. Moreover, it is also proven to achieve low diameter, degree and high resiliency and repair if any event of a take-down of a fraction of botnet node occurs. Anonymity is achieved through the periodically changing address of the bot during waiting stage. Unlike current botnets, secure communications is achieved by encryption in OnionBot through Tor and SSL. The threat environment will continue to grow due to its capability to offer new services such as botnet-for-rent and distribution computation platform for rent. By taking advantage of payment through Bitcoins, Silk Road 2.0 in Tor business operations can be carried out and botnets can be instructed to perform CPU intensive operations such as bitcoin mining or password cracking. Furthermore, the estimated threats due to OnionBots could be higher since they are robust to partitioning even when large fraction of these bots are taken down simultaneously. (Sanatinia and Noubir 2015) (Chaabane, Manils and Kaafar 2010)

\subsection{Threats due to ransomware}

According to (Cygnus Business Media Inc. 2015) ,threats due to ransomware are one of the biggest threats. In ransomware, the user machine is infected and files and applications are held hostage until a fee is paid. The threat is very high because ransomware is an automatic customer service-type model where the malware will install and then victim can pay through bitcoins. It doesn’t involve any human interaction for payment process.Cryptolocker and Cryptowall are two of the ransomwares mentioned in (Cygnus Business Media Inc. 2015).Furthermore, the report also states that malware economy is ever growing and ransomware and cryptocurrency such as bitcoins are helping attacker monetize their actions.

\section{Threat Assessment}

According to the survey of anonymity technologies in (Bingdong Li 2013), newer systems such as Tor, I2P are gaining popularity. Among the current anonymity systems, Tor networks have more number of volunteers than I2P. From previous attacks it is evident that, Botnet over Tor is growing and among all the privacy infrastructures threats due to subverting Tor will be highest. According to (Casenove and Miraglia 2014), even though botnets over Tor is good solution it is still not perfect. They do not provide ultimate resilience and even with Tor, a centralized botnet can have vulnerability such as single point of failure. When a botnet is integrated with the Tor infrastructure, a lot of attention is raised as it creates instability in a stable network such as Tor. Hence when a botnet comprising of millions of nodes joins the Tor network in a short span, it can be easily detected. Even from a client point of view, a botnet using Tor can leave traces. A malware runs Tor client as external process and if the client was not installed previously, exposure of malware activity would be trivial. By cross verifying the running processes, malware can be detected by identifying the Tor client process. By using such detection techniques, the estimated threat due to the tor-based-botnets can be greatly reduced. In case of OnionBot, the estimated level of threat is very high due to its robust nature and anonymity. By using the suggested mitigation technique Sybil Onion Attack Protocol (SOAP), the botnets can be neutralized. This might reduce the risk due to such bots but due to the continuous development of various variants of such botnets, threats due to these corresponding models also increases. Successful removal of such threats may not be guaranteed but measures to reduce the threat’s effectiveness and mitigation of risks due to such threats need to be addressed. 

\section{Conclusion}

In this paper, various kinds of privacy infrastructures are discussed and the rising threats of subverting the privacy infrastructure are analyzed.Since Tor is the most prominent privacy infrastructure, it is attracting more threats.Major threat is due to botnet over Tor infrastructure which has been the basis for prominent botnet attacks since 2010.Severity of threats with respect to new variations of botnets such as OnionBots are high due to its robustness and resiliency.Since research with respect to Onionbots propose effective mitigation techniques such as SOAP,risk due to the exploitation of these threats can be reduced but not completely avoided. Moreover, the severity of threats is still high. With the rapid development of various robust botnet variants, design mitigation strategies need to be implemented rapidly and effectively. Further research is recommended to achieve effectiveness in detection and mitigation of threats due to these infrastructures.


\begin{thebibliography}{99}

\bibitem{c1} ALSABAH, MASHAEL, and IAN GOLDBERG. "Performance and Security Improvements for Tor: A Survey." ACM, November 2016.
\bibitem{c2} Bingdong Li, Esra Erdin,Mehmet Hadi Gunes,George Bebis,Todd Shipley. "An overview of anonymity technology usage." ScienceDirect, 2013.
\bibitem{c3} Casenove, Matteo, and Armando Miraglia. "Botnet over Tor: The Illusion of Hiding." IEEE, 2014.
\bibitem{c4} Chaabane, Abdelberi, Pere Manils, and Mohamed Ali Kaafar. "Digging into Anonymous Traffic: a deep analysis of the Tor anonymizing network." IEEE, 2010.
\bibitem{c5} Chakravarty, Sambuddho. "Traffic Analysis Attacks and Defenses in Low Latency Anonymous Communication." PhD Thesis, Graduate School of Arts and Sciences, COLUMBIA UNIVERSITY, 2014.
\bibitem{c6} Constantin, Lucian. "Tor network used to command Skynet botnet." December 2012. http://www.techworld.com/ news/security/tor-network-used-command-skynet-botnet-3415592/. (accessed December 2016).
\bibitem{c7} Cygnus Business Media Inc. "Hacker tactics becoming much more sophisticated, report finds." ProQuest, 2015.

\bibitem{c8} D.Brown. "Resilient Botnet Command and Control with Tor." Defcon 18. 2010. http://www.defcon.org/images/defcon-18/dc- 18-presentations/D.Brown/DEFCON-18-Brown-TorCnC.pdf.
\bibitem{c9} Diaz, Jesus, David Arroyo, and Francisco B. Rodriguez. "Fair anonymity for the Tor network." arXiv, Dec 2014.

\bibitem{c10} Dunn, John. "Mevade botnet miscalculated effect on Tor network, says Damballa." September 2013. http://www.techworld. com/news/security/mevade-botnet-miscalculated-effect-on-tor-network-says-damballa-3468988/ (accessed December 2016).

\bibitem{c11} EDMAN, MATTHEW, and BU ̈LENT YENER. "On Anonymity in an Electronic Society: A Survey of Anonymous Communication Systems." ResearchGate, 2009.
\bibitem{c12} Erdin, Esra. "Anonymous Communication Systems: Usage Analysis and Attack Mechanisms." Thesis, Department of Computer Science, University of Nevada, Reno, 2012.
\bibitem{c13} Gottesman, Yotam. "RSA Uncovers New POS Malware Operation Stealing Payment Card \& Personal Information." January 2014. https://blogs.rsa.com/rsa-uncovers-new-pos-malware-operation-stealing-payment-card- personal-information/. (accessed December 2016).
\bibitem{c14} Groš, Stjepan, Marko Salkic ́, and Ivan Šipka. "Protecting TOR exit nodes from abuse." IEEE, May 2010.
\bibitem{c15} He, Gaofeng, Ming Yang, Junzhou Luo, and Xiaodan Gu. "A novel application classification attack against Tor ." IEEE, July 2015.

\bibitem{c16} Johnson, Beverly. "THE ADVANTAGES AND DISADVANTAGES OF THE DEEP WEB, TOR NETWORK, VIRTUAL CURRENCIES AND THE REGULATORY CHALLENGES THEREOF." Capstone Project, Master of Science in Economic Crime Management, Utica College, 2014.

\bibitem{c17} KATE, ANIKET, GREG M. ZAVERUCHA, and IAN GOLDBERG. "Pairing-Based Onion Routing with Improved Forward Secrecy." ACM, 2010.

\bibitem{c18} Khandhar, Pallav. "Banking Botnets Persist Despite Takedowns." SecureWorks,Inc. April 22, 2015. https://www.secureworks.com/research/banking-botnets-persist-despite-takedowns (accessed December 17, 2016).
\bibitem{c19} Sanatinia, Amirali, and Guevara Noubir. "OnionBots: Subverting Privacy Infrastructure for Cyber Attacks." IEEE, June 2015.
\bibitem{c20} ServiceObjects. "Tor: The good. The bad. The anonymous." ServiceObjects. December 2013.https://www.serviceobjects.com/resources/articles-whitepapers/tor-network-whitepaper (accessed December 2016).
\bibitem{c21} Syverson, Paul, and Griffin Boyce. "Genuine onion: Simple, Fast, Flexible, and Cheap Website Authentication." (ResearchGate) Junw 2015.

\bibitem{c22} Tails. Tails. 2016. https://tails.boum.org/doc/anonymous\_internet/i2p/index.en.html.
\bibitem{c22} Towards an Autonomous System Monitor for Mitigating Correlation Attacks in the Tor Network. PhD Thesis, Department of Social Informatics, Kyoto University, arXiV, 2016.








\end{thebibliography}
\end{document}